\documentclass[preprint,aps,amsmath,showpacs]{revtex4}
\usepackage{graphics}

\begin{document}

\title{Targeting qubit states using open-loop control}
\author{C. D'Helon}
\author{V. Protopopescu}
\email{protopopesva@ornl.gov}
\author{R. Perez}
\affiliation{Center for Engineering Science Advanced Research, Computer Science and Mathematics Division,\\ Oak Ridge National Laboratory, Oak Ridge, TN 37831-6355 USA}

\begin{abstract}
We present an open-loop (bang-bang) scheme which drives an open two-level quantum system to any target state, while maintaining quantum coherence throughout the process. The control is illustrated by a realistic simulation for both adiabatic and thermal decoherence. In the thermal decoherence regime, the control achieved by the proposed scheme is qualitatively similar, at the ensemble level, to the control realized by the quantum feedback scheme of Wang, Wiseman, and Milburn [Phys. Rev. A {\bf 64} 063810 (2001)] for the spontaneous emission of a two-level atom. The performance of the open-loop scheme compares favorably against the quantum feedback scheme with respect to robustness, target fidelity and transition times.
\end{abstract}

\pacs{03.67.Hk, 03.65.-w, 89.70.+c}
\maketitle

\newpage
\section{Introduction}
\label{Introduction}

Over the last few years, a number of open-loop (bang-bang) control schemes have been proposed to eliminate the effects of decoherence for a single two-level quantum system in contact with the environment, by using an external control \cite{Tombesi_ol,Viola98,Viola99a,Viola99b}. This type of control, pioneered by Lloyd and Viola (see Refs. \cite{Viola98,Viola99a,Viola99b} and references therein), relies on applying frequent control pulses to the two-level system in order to cancel the decoherence effects of the system-environment interaction. Vitali and Tombesi \cite{Tombesi_ol} have also considered applying a sequence of frequent parity kicks, as well as an appropriate stochastic modulation, to achieve the same goal. It has been shown that, in either approach, the decoherence of the system can be effectively suppressed if the pulse rate is much higher than the decoherence rate due to the system-environment interaction. In other words, a fast rate control "freezes" decoherence in a manner similar to the quantum Zeno effect. 


Alternative approaches have also been considered to eliminate or mitigate the undesirable effects of decoherence in open quantum systems, e.g., decoherence free subspaces (DFS) \cite{Zanardi97,Lidar98}, quantum error correction \cite{Preskill98,Knill00,NC} and quantum feedback \cite{Tombesi_qf,Wang01}.

Inspired by Lloyd and Viola's seminal results, we have recently proposed a new implementation of open-loop quantum control \cite{Protopopescu02}, which tailors control pulses more efficiently, by taking into account the possible knowledge of the decoherence function. Several new features of the scheme are: (i) decoherence and control act simultaneously within a realistic model that includes them from first principles; (ii) the required control is directly related to and calculated from the decoherence effects, which presents the practical advantage of maintaining the frequency and amplitude of the required controls at minimal levels; and (iii) the effect of imperfect pulses on the efficiency of the control was assessed and found to be quite tolerable, even for rather large noise amplitudes. In this paper, our aim is to minimize the effect of decoherence due to interactions between the quantum system and its environment, while concommitantly driving the system from an initial state to an arbitrary target state.

Recently, Khaneja \textit{et al.} \cite{Khaneja_control} used optimal control to steer a quantum system to a target state in a minimum time, while assuming that the effects of relaxation are negligible over that timescale. They have also demonstrated that controllability for a closed quantum system is equivalent to the universality that is demanded from a quantum computer. In follow-on work, Khaneja \textit{et al.} \cite{Khaneja_relax} used optimal control to minimize the difference between the actual density matrix at the final time and the target density matrix, for two spins in the presence of relaxation. We note that their relaxation model relies on \textit{phenomenological} decay rates to describe general decoherence processes, without any explicit reference to the environment, and the control is applied \textit{a posteriori}, to the ``reduced'' equations. This is in contrast with the present scheme, where the interaction with the environment and the control are accounted for from first principles, in the full Hamiltonian description of the system.

We illustrate our approach for pure states of a two-level system, i.e., a qubit, which is in contact with the environment. Both the initial and the target states are known \textit{a priori}, and the unitary evolution of the two-level system is driven by frequent control pulses that are customized, if the decoherence function is known. This is the case - at least in principle - if one has an explicit model for the environment and its interaction with the qubit. 

Using numerical simulations, we show explicitly that the qubit can be driven between any two points (initial and target states) on the curve determined by the intersection of either the $x-z$, or the $y-z$ planes with the Bloch sphere. In general, the state of the qubit can be driven between any two points on the surface of the Bloch sphere by using a sequence of two different control Hamiltonians. 

In one particular instance, namely for thermal decoherence and external control proportional to the $\sigma_y$ polarization, the system we consider is formally similar to the system used in the quantum feedback scheme proposed by Wang {\it et al.} \cite{Wang01}, which models the evolution of a two-level atom that emits spontaneously, using stochastic quantum trajectories. Photocurrent feedback from perfect homodyne detection of spontaneous emission is used to alter the atom-environment dynamics, to drive the atom to a pure target state. Of course, the feedback delay time has to be shorter than the spontaneous emission timescale, in order to successfully maintain the atom in the target state.

Whenever warranted, we shall discuss the performance of the proposed open-loop control in comparison with the performance of quantum feedback control, in order to understand the limitations of each scheme, and characterize the regimes in which they work best. In so doing, we also generalize the original feedback control scheme propsed by Wang {\it et al.}, by using different control Hamiltonians.

The remainder of the paper is organized as follows: In Section 2, we give an overview of the spin-boson model used in our targeting control scheme, and in Section 3 we outline the control strategy employed to achieve a particular target state. In Section 4, we briefly review the quantum feedback scheme proposed by Wang {\it et al.} \cite{Wang01} for a spontaneously-emitting two-level atom. Finally, in Section 5, we present the results of our numerical simulations, and discuss them in comparison with the results of the quantum feedback scheme.

\section{The Model}
\label{The Model}
The spin-boson model \cite{Weiss99} is often used to describe a two-level system $\{|1>,|2>\}$ interacting with a large number of boson modes representing the environment. The complete Hamiltonian \cite{Protopopescu02},
\begin{equation}
H = H_s + H_e + H_i + H_c,
\end{equation}
accounts for the two-level system, the environment, the interaction between the two, and an external control. The Hamiltonian of the two-level system is
\begin{equation}
H_s = \sum_{i=1}^2P_{ii}E_i,
\end{equation}
where $P_{ii}$ denote the projection operators $P_{ii}=|i\rangle \langle i|$, and $E_i$ are the corresponding energies. We take $E_{1}=-\hbar\omega_0/2$ and $E_{2}=\hbar\omega_0/2$, so that $H_s = \hbar\omega_0\sigma_z $. The Hamiltonian of the environment is
\begin{equation}
H_e = \hbar\sum_{q=1}^{\infty}\omega_{0q}a^\dagger_q a_q,
\end{equation}
where $a_q$, $a^\dagger_q$ are the annihilation and creation operators respectively, for the $q$-th boson mode, and $\omega_{0q}$ is the corresponding frequency.

The interaction between the two-level system and the environment, 
\begin{equation}
H_i = -\hbar\epsilon(\alpha_x\sigma_x + \beta_z\sigma_z)\sum_{q=1}^{\infty}(\Omega_q^\star a_q + \Omega_q a^\dagger_q)
\end{equation}
is responsible for the quantum decoherence. Adiabatic decoherence ($\alpha_x=0$, $\beta_z=1$) acts on a short timescale, leading to the decay of the off-diagonal density matrix elements. On a longer timescale, thermal decoherence ($\alpha_x=1$, $\beta_z=0$) changes all of the density matrix elements, and leads to the exponential decrease of the excited state population of the two-level system. These two timescales can be accounted for by adjusting the relative magnitude of the coefficients, $\alpha_x$ and $\beta_z$. The interaction with the environment is parametrized by $\epsilon$ (whose magnitude indicates the strength of the coupling) and can result in a phase flip or a bit flip (a bit-phase flip can also be obtained either by combining the two effects or by including a $\sigma_y$ term).  However, from a formal viewpoint, the effects of $\sigma_x$ and $\sigma_y$ in the \textit{interaction} Hamiltonian are essentially similar, and we shall not consider the latter as far as $H_i$ is concerned.

In general, the control Hamiltonian is proportional to a linear superposition of the pseudo-spin operators $\sigma_x$ and $\sigma_y$,
\begin{equation}
H_c = -\hbar\Omega_F V(t) [c_x\sigma_x + c_y\sigma_y],
\end{equation}
which can be specified by adjusting the phase of the driving field. The control is a classical coherent driving field, which is applied as a series of frequent pulses, $V(t)$. We assume that these pulses act strongly upon the two-level system, via its polarization, but do not affect the environment significantly.

The dynamics of the two-level system is expressed in terms of parameters normalized to the Rabi frequency, $\Omega_F$. The time evolution is described in terms of dimensionless Rabi time units, $\tau=\Omega_F t$, and all the frequencies, namely $\omega_q=\omega_{oq}/\Omega_F$, $\omega=\omega_o/\Omega_F$, $g_q=\Omega_q/\Omega_F$ and $\omega_{12}=\Omega_{12}/\Omega_F$, are also renormalized with respect to the Rabi frequency. The frequency $\Omega_{12}$ is the transition frequency between the energy levels of the two-level system.

To simplify the calculations and render them more transparent, it is convenient to write the evolution of the system in the interaction representation, and consider the adiabatic and thermal decoherence regimes separately. The unitary evolution operator in the interaction representation, $U_I$, is related to the unitary evolution operator in the Schr\"odinger representation, $U$, by

\begin{equation}
U_I(s,e,\tau)=e^{i H_0 \tau /\hbar} U(s,e,\tau)
\end{equation}
where $H_0=H_s+H_e$.

In this representation, the evolution of the density matrix of the whole system (qubit plus environment) reads
\begin{equation}
\frac{\partial\rho}{\partial\tau}=-\frac{i}{\hbar}[H_I,\rho]
\end{equation}
where the interaction Hamiltonian contains both the interaction with the environment and the applied control:
\begin{equation}
H_I=H_{Ic}+H_{Ii},
\label{H_I}
\end{equation}
\begin{equation}
H_{Ii}=
\begin{cases}
-\epsilon\hbar\sigma_z \sum_{q}(g_q^\star a_q e^{-i\omega_q \tau} + g_q a^\dagger_q e^{i\omega_q \tau}) \\
-\epsilon\hbar [ \frac{\sigma_x + i \sigma_y}{2} \sum_q g_q^\star a_q e^{-i(\omega_{12}-\omega_q)\tau} + \frac{\sigma_x - i \sigma_y}{2} \sum_q g_q a^\dagger_q e^{i(\omega_{12}-\omega_q)\tau}) ]
\end{cases}
\end{equation}
for the adiabatic and thermal decoherence regimes respectively, and
\begin{equation}
H_{Ic}=-\frac{\hbar}{2}V(\tau)[C_x \sigma_x + C_y \sigma_y],
\label{H_Ic}
\end{equation}
where
\begin{eqnarray}
C_x &=& c_x\cos(\omega_0 t) + c_y\sin(\omega_0 t) \\
C_y &=& -c_x\sin(\omega_0 t) + c_y\cos(\omega_0 t).
\end{eqnarray}
These interaction-picture Hamiltonians have been calculated by using the rotating wave approximation, and assuming zero detuning, $\delta=\omega_{12}-\omega=0$.

The evolution equation for $\rho$ admits the formal solution 
\begin{equation}
\rho(s,e,\tau)=U_I(s,e,\tau)\rho(s,e,0)U_I^{\dagger}(s,e,\tau)
\end{equation}
where the evolution operator satisfies the equation
\begin{equation}
\frac{dU_I(s,e,\tau)}{d\tau}=-\frac{i}{\hbar}H_I U_I(s,e,\tau)
\label{evolution_eqn_U_I}
\end{equation}
given the initial condition $U_I(s,e,0)=1$. The formal solution of Eq.(\ref{evolution_eqn_U_I}) is \cite{Scully_qo},
\begin{equation}
U_I(s,e,\tau)={\mathcal T}[\exp\{-\frac{i}{\hbar}\int_0^\tau d\tau^\prime H_I(\tau^\prime)\}],
\end{equation}
where ${\mathcal T[\hspace{5pt}]}$ represents the time-ordering operator. The state of the system at $\tau = 0$ is assumed to be described by the density matrix:
\begin{equation}
\rho(s,e,0)=\rho(s,0)\otimes\rho(e,0)
\end{equation}
where the environment is in thermal equilibrium,
\begin{equation}
\rho(e,0)=\prod_q{\rho_q(e,0)}
\end{equation}
\begin{equation}
\rho_q(e,0)=[1-e^{-\frac{\hbar\omega_{oq}}{kT}}]^{-1}\sum_{n_q}e^{-\frac{\hbar\omega_{oq}}{kT}n_q}|n_q\rangle\langle n_q|
\end{equation}
and
\begin{equation}
\rho(s,0)=\sum_{i,j=1}^2 \rho_{ij}(0)P_{ij}.
\end{equation}

Since the various terms in the interaction Hamiltonian $H_I$ do not commute, the complete evolution operator, $U_I(s,e,\tau)$, cannot be calculated exactly, except in very special cases. By writing the interaction Hamiltonian as a sum of two non-commuting operators, $H_I=H_1+H_2$, we can expand the evolution operator into an infinite product of exponentials,
\begin{equation}
U_I(s,e,\tau)=e^{-\frac{i}{\hbar}\int_0^\tau dt H_1(t)} \times e^{-\frac{i}{\hbar}\int_0^\tau dt H_2(t)} \times e^{-(\frac{i}{\hbar})^2\int_0^\tau dt \int_0^t dt^\prime [H_1(t),H_2(t^\prime)]} \times \dots
\label{U_I_full}
\end{equation}
using the general Baker-Hausdorff theorem \cite{Gardiner_qn}. Assuming that the incremental effects of the control and interaction Hamiltonians are relatively small, we approximate the evolution operator to the first order in the magnitude of the control pulses, $V(\tau)$, and the coupling strength parameter, $\epsilon$, of the system-environment interaction.

We consider the evolution operator, $U_I(s,e,\tau)$, and the density matrix elements for the adiabatic and thermal decoherence regimes, separately. The latter case describes population changes in the two-level system, and encompasses the evolution of a spontaneously-emitting atom, in the quantum feedback scheme proposed by Wang et al. \cite{Wang01}.

\subsection{Adiabatic Decoherence Regime}
The adiabatic decoherence regime describes phase decay of the two-level system. For the sake of simplicity, we assume that the control Hamiltonian is proportional to $\sigma_x$, i.e., $(C_x = 1,C_y = 0)$,
\begin{equation}
H_{Ic}=-\frac{\hbar}{2}V(\tau)\sigma_x,
\end{equation}
and we identify the control Hamiltonian, $H_{Ic}$, with $H_1$, and the system-environment interaction Hamiltonian, $H_{Ii}$, with $H_2$. Therefore, by neglecting the commutator terms in Eq. (\ref{U_I_full}) the evolution operator can be approximated by
\begin{equation}
U_I(s,e,\tau) \approx exp\{-\frac{i}{\hbar}\int_0^\tau dt H_{Ic}(t)\} \times exp\{-\frac{i}{\hbar}\int_0^\tau dt H_{Ii}(t)\}.
\end{equation}
This approximation is correct to first order in the magnitude of the control and interaction Hamiltonians.

Performing the time integrals in the exponents, we obtain
\begin{equation}
U_I(s,e,\tau)=
\exp\{i \sigma_x I(\tau)\}\times\exp\{-2 \sigma_z Q_-(\tau)\}
\label{U_I_a}
\end{equation}
where
\begin{equation}
I(\tau)=\frac{1}{2}\int_0^\tau d\tau^\prime V(\tau^\prime)
\label{I_tau}
\end{equation}
\begin{equation}
Q_-(\tau)=\frac{\epsilon}{2}\sum_q (M_q a^\dagger_q - M_q^\star a_q)
\label{Q-_tau}
\end{equation}
and
\begin{equation}
M_q(\tau)=
\frac{g_q}{\omega_q}(1-e^{i\omega_q\tau}).
\end{equation}

We obtain the following approximate expressions for the evolution operator and its adjoint:
\begin{equation}
U_I(s,e,\tau)=E_0+iE_x\sigma_x+E_y\sigma_y+E_z\sigma_z
\end{equation}
and
\begin{equation}
U_I^{\dagger}(s,e,\tau)=E_0-iE_x\sigma_x-E_y\sigma_y-E_z\sigma_z
\end{equation}
where
\begin{eqnarray}
E_0 &=& \cosh(2Q_-(\tau))\cos(I(\tau)) \\
E_x &=& \cosh(2Q_-(\tau))\sin(I(\tau)) \\
E_y &=& -\sinh(2Q_-(\tau))\sin(I(\tau)) \\
E_z &=& -\sinh(2Q_-(\tau))\cos(I(\tau)).
\end{eqnarray}

The density matrix elements of the reduced two-level system can then be solved explicitly by tracing out the environmental modes, 
\begin{equation}
\rho_{ij}(s,\tau)=Tr_e\{\sum_{k,l=1}^2 \langle i|U_I P_{kl} \rho(e,0) U_I^{-1} |j\rangle \rho_{kl}(0)\}
\label{rho_ij}
\end{equation}
for $i,j=1,2$, assuming that the environment is initially in a thermal state. We obtain:

\begin{eqnarray}
\rho_{11}&=&\rho_{11}(0)\cos^2 I + \rho_{22}(0)\sin^2 I - i[\rho_{12}(0)-\rho_{21}(0)]e^{-g_{ad}}\cos I\sin I,
\label{rho_11_a} 
\\
\rho_{22}&=&\rho_{22}(0)\cos^2 I + \rho_{11}(0)\sin^2 I + i[\rho_{12}(0)-\rho_{21}(0)]e^{-g_{ad}}\cos I\sin I,
\label{rho_22_a} 
\\
\rho_{12}&=&\rho_{12}(0)e^{-g_{ad}}\cos^2 I + \rho_{21}(0)e^{-g_{ad}}\sin^2 I + i (\rho_{22}(0) - \rho_{11}(0))\cos I\sin I,
\label{rho_12_a} 
\\
\rho_{21}&=&\rho_{21}(0)e^{-g_{ad}}\cos^2 I + \rho_{12}(0)e^{-g_{ad}}\sin^2 I - i (\rho_{22}(0) - \rho_{11}(0))\cos I\sin I,
\label{rho_21_a}
\end{eqnarray}
\noindent
where $I=I(\tau)$, and 
\begin{equation}
g_{ad} := g_{ad}(\tau)=\gamma\int_0^\infty d\omega G(\omega)(1-\cos{\omega \tau})\coth\frac{\beta_0\omega}{2}
\end{equation}
\noindent
is the adiabatic decoherence function obtained by Palma \cite{Palma96}. The dimensionless constant $\gamma$ depends on the dipole moment of the two-level system and on the Rabi frequency.

\subsection{Thermal Decoherence Regime}
The thermal decoherence regime describes population changes in the two-level system, and encompasses the evolution of a spontaneously-emitting atom. We use the general form of the control Hamiltonian in Eq.(\ref{H_Ic}), and we decompose the interaction Hamiltonian $H_I$ into two operators, $H_1$ and $H_2$, which are proportional to $\sigma_x$ and $\sigma_y$, respectively. 

By neglecting the commutator terms in Eq. (\ref{U_I_full}), the evolution operator can be approximated by
\begin{equation}
U_I(s,e,\tau) \approx exp\{-\frac{i}{\hbar}\int_0^\tau dt (H_{Ic_x}(t)+H_{Ii_x}(t))\} \times exp\{-\frac{i}{\hbar}\int_0^\tau dt (H_{Ic_y}(t)+H_{Ii_y}(t))\},
\end{equation}
where $H_{Ic_x},H_{Ii_x}$ and $H_{Ic_y},H_{Ii_y}$ correspond to the terms in the control Hamiltonian $H_{Ic}$ and the system-environment interaction Hamiltonian $H_{Ii}$, which are proportional to $\sigma_x$ and $\sigma_y$, respectively. This approximation is correct to first order in the magnitude of the control and interaction Hamiltonians.

Performing the time integrals in the exponents, we get
\begin{equation}
U_I(s,e,\tau)=
\exp\{i C_x \sigma_x I(\tau) - \sigma_x Q_-(\tau)\}\times\exp\{i C_y \sigma_y I(\tau) + i \sigma_y Q_+(\tau)\}
\label{U_I_t}
\end{equation}
where
\begin{equation}
Q_+(\tau)=\frac{\epsilon}{2}\sum_q (M_q a^\dagger_q + M_q^\star a_q),
\label{Q+_tau}
\end{equation}
\begin{equation}
M_q(\tau)=\frac{g_q}{\omega_{12}-\omega_q}(1-e^{i(\omega_{12}-\omega_q)\tau}),
\end{equation}
and $I(\tau)$, $Q_-(\tau)$ are defined by Eqs. (\ref{I_tau})-(\ref{Q-_tau}).

We obtain the following approximate expressions for the evolution operator and its inverse:
\begin{equation}
U_I(s,e,\tau)=E_0+iE_x\sigma_x+E_y\sigma_y+E_z\sigma_z
\end{equation}
and
\begin{equation}
U_I^{\dagger}(s,e,\tau)=E_0-iE_x\sigma_x-iE_y\sigma_y+iE_z\sigma_z
\end{equation}
where
\begin{eqnarray}
E_0 &=& \cos(C_x I(\tau)+iQ_-(\tau))\cos(C_y I(\tau)+Q_+(\tau)) \\
E_x &=& \sin(C_x I(\tau)+iQ_-(\tau))\cos(C_y I(\tau)+Q_+(\tau)) \\
E_y &=& \cos(C_x I(\tau)+iQ_-(\tau))\sin(C_y I(\tau)+Q_+(\tau)) \\
E_z &=& \sin(C_x I(\tau)+iQ_-(\tau))\sin(C_y I(\tau)+Q_+(\tau)).
\end{eqnarray}
To calculate $U_I^{\dagger}(s,e,\tau)$, we have used the fact that $Q_-(\tau)$ and $Q_+(\tau)$ commute in the first order approximation of the coupling strength parameter, $\epsilon$.

The environment modes are then traced out to obtain the elements of the reduced density matrix, in the same way as for the adiabatic case (see Eq.(\ref{rho_ij})), assuming as before, that the environment is initially in a thermal state:
\begin{eqnarray}
\rho_{11} &=& \frac{1}{2} + \frac{1}{2}(\rho_{11}(0)-\rho_{22}(0))e^{-2g_{th}}\cos(2C_xI)\cos(2C_yI) \nonumber \\
& & - Re\{\rho_{12}(0)\}e^{-2g_{th}}\cos(2C_xI)\sin(2C_yI) - iIm\{\rho_{12}(0)\}e^{-g_{th}}\sin(2C_xI)
\label{rho_11_t} 
\\
\rho_{12} &=& [Re\{\rho_{12}(0)\}\cos(2C_yI)+Im\{\rho_{12}(0)\}\cos(2C_xI)]e^{-g_{th}} \nonumber 
\\
& & + i Re\{\rho_{12}(0)\}\sin(2C_xI)\sin(2C_yI)e^{-2g_{th}} \nonumber \\
& & + \frac{1}{2}(\rho_{11}(0)-\rho_{22}(0))[e^{-g_{th}}\sin(2C_yI) - i e^{-2g_{th}}\sin(2C_xI)\cos(2C_yI)]
\label{rho_12_t} 
\\
\rho_{21} &=& [Re\{\rho_{21}(0)\}\cos(2C_yI)+Im\{\rho_{21}(0)\}\cos(2C_xI)]e^{-g_{th}} \nonumber \\
& & - i Re\{\rho_{21}(0)\}\sin(2C_xI)\sin(2C_yI)e^{-2g_{th}} \nonumber \\
& & + \frac{1}{2}(\rho_{11}(0)-\rho_{22}(0))[e^{-g_{th}}\sin(2C_yI) + i e^{-2g_{th}}\sin(2C_xI)\cos(2C_yI)]
 \\
\rho_{22} &=& \frac{1}{2} - \frac{1}{2}(\rho_{11}(0)-\rho_{22}(0))e^{-2g_{th}}\cos(2C_xI)\cos(2C_yI) \nonumber \\
& & + Re\{\rho_{12}(0)\}e^{-2g_{th}}\cos(2C_xI)\sin(2C_yI) + iIm\{\rho_{12}(0)\}e^{-g_{th}}\sin(2C_xI)
\label{rho_22_t}
\end{eqnarray}

\noindent
where the thermal decoherence function, $g_{th}$, is given by 
\begin{equation}
g_{th} := g_{th}(\tau)=\gamma\int_0^\infty d\omega \frac{1-cos[(\omega_{12}-\omega) \tau]}{(\omega_{12}-\omega)^2}\omega^3 \coth(\beta_0\omega/2)\exp(-\omega/\omega_c).
\end{equation}

\section{Control Strategy}
\label{Control Strategy}

The control strategy in Ref. \cite{Protopopescu02} was to continually adjust the control pulses in a customized fashion to counter the effects of the decoherence and maintain the density matrix elements unchanged. In this paper, the goal is to drive the qubit to the final target state while, at the same time, maintaining quantum coherence throughout the process. The reduced density matrix of the qubit depends on the known decoherence function and the applied control, which is determined by equating each element of the actual reduced density matrix, $\rho_{ij}$, in turn, with the corresponding element of the density matrix for the target state.

Since in general, a large jump in the state of the qubit cannot be achieved in one step,the qubit is driven between the initial and target states via a number of intermediate states. This ensures controllability i.e., the existence of feasible solutions for the proposed control.


Each control cycle contains 8 control steps, since there are 8 real equations to solve, for the reduced density matrix of the two-level system. The sequence of eight real transcendental equations is solved in a prescribed order. While this particular order is not essential, we shall use it consistenly here, as shown below, to make it easier to follow our strategy. The algorithm is applied identically for either the adiabatic or the thermal case, thus we describe it for a generic decoherence function, denoted $g$.


The intermediate states spanning the trajectory between the initial and target values can be chosen explicitly, the simplest way being to interpolate linearly between the initial and target states, or alternatively, they can be selected by minimizing the absolute difference between the initial and target values for a restricted range of control pulses. 


In the first cycle of eight time steps, we start by driving the first component $\rho_{11R}(0)$ to a new intermediate value, $\zeta_{11R}^{(1)}$, using the control pulse $I(1)$, which is a solution of the equation $\rho_{11R}(1)=\rho_{11R}(g(1),I(1))=\zeta_{11R}^{(1)}$. After the first time step, all the other components have been modified by the effect of the decoherence, $g(1)$, and the effect of the control pulse, $I(1)$, so the values of the matrix elements are given by:

\begin{eqnarray}
\rho_{11R}(1) &=& \rho_{11R}(g(1),I(1)) = \zeta_{11R}^{(1)} \\
\rho_{11I}(1) &=& \rho_{11I}(g(1),I(1)) \\
\rho_{12R}(1) &=& \rho_{12R}(g(1),I(1)) \\
\rho_{12I}(1) &=& \rho_{12I}(g(1),I(1)) \\
\rho_{21R}(1) &=& \rho_{21R}(g(1),I(1)) \\
\rho_{21I}(1) &=& \rho_{21I}(g(1),I(1)) \\
\rho_{22R}(1) &=& \rho_{22R}(g(1),I(1)) \\
\rho_{22I}(1) &=& \rho_{22I}(g(1),I(1))
\end{eqnarray}

During the next seven time steps, the control strategy is implemented in a similar fashion. For example, at the second time step, we consider the equation for $\rho_{11I}(2)=\rho_{11I}(g(2),I(1)+I(2))$. For relatively short time steps, and by the semigroup property, this is also equal to $\zeta_{11I}^{(1)}(g(1),I(2))$. To determine the control pulse required to set $\rho_{11I}(2)$ equal to a new value $\zeta_{11I}^{(1)}$, we find the control $I(2)$ which solves the equation $\rho_{11I}(2) = \zeta_{11I}^{(1)}$. After applying this control, the values of the matrix elements at the second time step are given by:

\begin{equation}
\rho_{ijR/I}(2) = \rho_{ijR/I}(g(2),I(1)+I(2)).
\end{equation}

By the end of the first cycle of eight time steps, the state has shifted slightly towards the first intermediate state, $\zeta^{(1)}$. The final values of the eight components at the end of the first cycle become the initial values for a second cycle of eight time steps, which then aims to drive the qubit to the second intermediate state, $\zeta^{(2)}$, using the same method.

This procedure is repeated until the final target state is reached, and is then maintained for as long as required. The total transition time to drive the qubit from the initial to the target state depends on the number of intermediate states and the frequency of the control pulses. After the initial transient, the controls stabilize and the whole control cycle repeats periodically.

\section{Quantum Feedback Control}
\label{Quantum Feedback Control}

In this section, we shall briefly review the quantum feedback scheme proposed by Wang \textit{et al.} \cite{Wang01}. The theory of quantum feedback has been developed in the last ten years \cite{Wiseman_feedback}, and recently there has been growing interest in using it to cancel the effects of decoherence. This can be achieved by using the continuous measurement record resulting from the coupling of a system to its environment, to cancel out the effect of the interaction with the environment.

The system considered by Wang \textit{et al.} \cite{Wang01} is a two-level atom, which is driven by a resonant classical field of strength $\alpha\in(-\infty,\infty)$. The lowering operator is denoted by $\sigma=|1\rangle\langle 2|$; its adjoint $\sigma^\dagger=|2\rangle\langle 1|$ is the raising operator, and performs the inverse operation. The environment is represented by electromagnetic modes into which the atom emits spontaneously at a decay rate $\gamma$. The reduced master equation describing this system, with no feedback, is derived from the interaction Hamiltonian \cite{Milburn_qo}:
\begin{equation}
H_I = \hbar \alpha \sigma_y + \hbar \{ \sigma^\dagger\sum_q g_q^* a_q + \sigma\sum_q g_q a_q^\dagger \},
\label{qf_H_I}
\end{equation}
which is identical to the interaction Hamiltonian considered in Section 2.2 for the thermal decoherence regime. Phase decay of the two-level atom, due to atomic collisions or other processes, has been ignored in the quantum feedback scheme, i.e. there is no $\sigma_z$-coupling for the system-environment interaction. 
The reduced master equation is given by
\begin{equation}
\dot{\rho} = -i\alpha[\sigma_y,\rho] + \mathcal{D}[\sqrt{\gamma}\sigma] \rho,
\label{me_nofeedback}
\end{equation}
where the Lindblad superoperator $\mathcal{D}$ is
\begin{equation}
\mathcal{D}[A]B = ABA^\dagger - \{A^\dagger A,B \}/2.
\end{equation}
We note that the same reduced master equation can be obtained for the spin-boson model we presented in Section 2, in the thermal decoherence regime, using a control Hamiltonian proportional to $\sigma_y$.

The effect of the classical driving field $\alpha$ is to rotate the state of the atom around the $y$-axis, in Bloch space. The Bloch space coordinates are defined by the vector $(x,y,z)$, related to the state density matrix by:
\begin{equation}
\rho = \frac{1}{2} (I + x\sigma_x + y\sigma_y + z\sigma_z).
\end{equation}
The distance from the centre of the Bloch sphere, $r=\sqrt{x^2+y^2+z^2}$, measures the purity of the state, with $r=1$ for a pure state, and $r=0$ for a maximally-mixed state. The stationary solutions for the driven two-level atom,
\begin{eqnarray}
x_{ss} &=& \frac{4\alpha\gamma}{\gamma^2+8\alpha^2} \\
y_{ss} &=& 0 \\
z_{ss} &=& \frac{-\gamma^2}{\gamma^2+8\alpha^2},
\end{eqnarray}
are limited to the lower half of the Bloch sphere $(z<0)$, in the $x-z$ plane, and in general have low purity, as shown in Fig. 2 of Ref. \cite{Wang01}. The ground state with no driving is the only stationary solution that is a pure state. We note that in general, the interaction Hamiltonian due to the classical driving field can be taken as a linear superposition of $\sigma_x$ and $\sigma_y$, as we have done for our control Hamiltonian $H_{Ic}$. Experimentally this superposition of polarizations can be adjusted by controlling the phase of the classical driving field.


In their scheme, Wang {\it et al.} assume that \textit{all} of the fluorescence from the atom is collected and used in a homodyne detection setup (see Fig. 1 in Ref. \cite{Wang01}), to measure the interference between the system and a local oscillator. In the ideal limit of large local oscillator amplitude, the point process of photocounts is changed into a continuous photocurrent with white noise. The interference is measured by the (normalized) difference between the recorded mean photocurrents,
\begin{equation}
\Delta I(t) = \sqrt{\gamma}\langle\sigma_x\rangle_c(t) + \zeta(t),
\end{equation}
given that the phase of the local oscillator is set to zero. The subscript $c$ means conditioned, as system averages are conditioned on the previous photocurrent record. The last term, $\zeta(t)$, represents Gaussian white noise so that,
\begin{equation}
\zeta(t) dt = dW(t),
\end{equation}
an infinitesimal Wiener increment defined by
\begin{eqnarray}
\{dW(t)\}^2 &=& dt, \\
E[dW(t)] &=& 0.
\end{eqnarray}

The conditioning process can be made explicit by calculating the changes of the system state in response to the measured photocurrent, using a stochastic Schr\"{o}dinger equation,
\begin{equation}
d|\psi_c(t)\rangle = \hat{A}_c(t)|\psi_c(t)\rangle dt + \hat{B}_c(t)|\psi_c(t)\rangle dW(t),
\end{equation}
which describes possible quantum trajectories, assuming that the initial state is pure at some point in time. The operator $\hat{A}_c(t)$ for the drift term, and the operator $\hat{B}_c(t)$ for the diffusion term are both conditioned, as they depend on the system average
\begin{equation}
\langle\sigma_x\rangle_c(t) = \langle\psi_c(t)|\sigma_x|\psi_c(t)\rangle.
\end{equation}
Of course, from the point of view of ensemble averages, the two-level atom still obeys the same master equation given above in Eq. (\ref{me_nofeedback}).

The difference between the recorded mean photocurrents, $\Delta I(t)$, is used to control the dynamics of the state in the $x-z$ plane. The atom-environment dynamics is altered by adding a photocurrent feedback, $\lambda \Delta I(t)$, to the amplitude of the classical driving field, using an electro-optic amplitude modulator. The effect of this feedback term is to change the drift, $\hat{A}_c(t)$, and the diffusion, $\hat{B}_c(t)$, operators, such that the feedback-modified master equation describing the quantum feedback scheme becomes:
\begin{equation}
\dot{\rho} = -i\alpha[\sigma_y,\rho] + \mathcal{D}[\sqrt{\gamma}\sigma-i\lambda\sigma_y] \rho.
\label{me_feedback}
\end{equation}

The two-level atom can be driven to, and subsequently maintained in an arbitrary pure target state $\hat{|\theta\rangle}$, in the $x-z$ plane,
\begin{equation}
\hat{|\theta\rangle} = \cos\biggl(\frac{\theta}{2}\biggr)|2\rangle + \sin\biggl(\frac{\theta}{2}\biggr)|1\rangle,
\end{equation}
by ensuring that only the phase of $\hat{|\theta\rangle}$ changes under this evolution:
\begin{equation}
\hat{|\theta\rangle} + d\hat{|\theta\rangle} = e^{id\chi}\hat{|\theta\rangle}.
\end{equation}
In terms of the stochastic Schr\"{o}dinger equation, this implies that the deterministic (drift) and noise (diffusion) terms are separately proportional to $\hat{|\theta\rangle}$,
\begin{eqnarray}
\hat{A}_c(t) dt \hat{|\theta\rangle} & \propto & \hat{|\theta\rangle} \\
\hat{B}_c(t) dW(t)] \hat{|\theta\rangle} & \propto & \hat{|\theta\rangle},
\end{eqnarray}
which can be achieved by setting the values of the driving and feedback parameters to:
\begin{eqnarray}
\alpha &=& \gamma/4\sin\theta\cos\theta, \\
\lambda &=& -\frac{\sqrt{\gamma}}{2}(1+\cos\theta).
\end{eqnarray}

A stability analysis of the scheme shows that the pure state $\hat{|\theta\rangle}$ can be achieved for all $\theta$, with the exception of $\theta=\pm\pi/2$, which correspond to equal superposition states on the equator of the Bloch sphere. The driving and feedback parameters for these special values are degenerate, thus the state is driven to a mixture of $\hat{|\frac{\pi}{2}\rangle}$ and $\hat{|-\frac{\pi}{2}\rangle}$.

The quantum feedback scheme described above relies on two major assumptions, that are at present impossible to achieve experimentally, namely: (i) perfect detection of spontaneous emission, and (ii) zero feedback delay time.

As the efficiency of the homodyne detection of resonance fluorescence from the two-level atom is decreased from perfect efficiency, the performance of the scheme degrades very quickly, except for target states that are close to the ground state. In general, detection with non-unit efficiency gives a stationary solution that represents a mixed state, rather than the desired pure target state. Wang \textit{et al.} have shown that the deviation from a pure target state is noticeable even for relatively high $(95\%)$ efficiencies, especially for states close to the equal superposition states at the equator of the Bloch sphere, which are inherently unstable.

Similarly, in order to successfully achieve a pure target state, the feedback delay time $\tau$ has to be much shorter than the time scale corresponding to the spontaneous emission rate. This limit corresponds to the Markovian regime, which allows the evolution equation to be written in the form of a master equation. If the feedback delay time becomes too large, $\tau \geq 0.02\gamma^{-1}$, the performance of the quantum feedback scheme suffers a degradation which is qualitatively similar to that observed for inefficient detection. For example, if we try to stabilize the atom in the excited state, in the non-Markovian regime, the purity of the average state is given by $r \approx 1-4\gamma\tau$. Of course, the deviation from a pure target state is much more noticeable for states close to the equal superposition states at the equator of the Bloch sphere.

\begin{figure}[tp]
(a)\scalebox{0.4}{\includegraphics{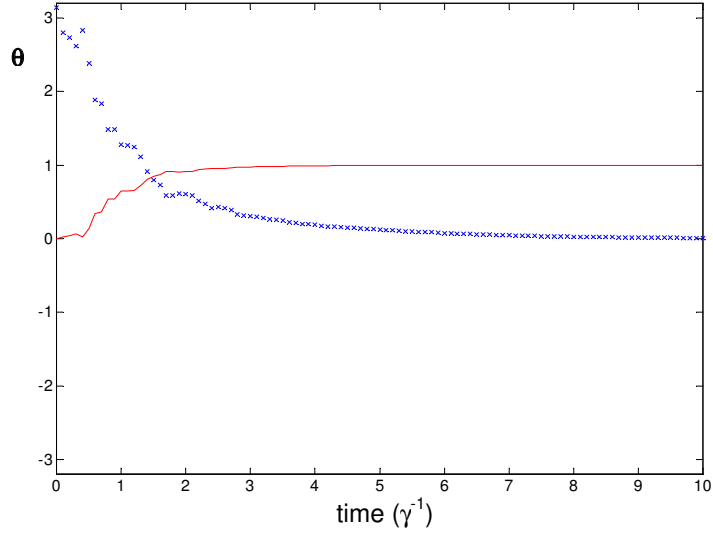}}
(b)\scalebox{0.4}{\includegraphics{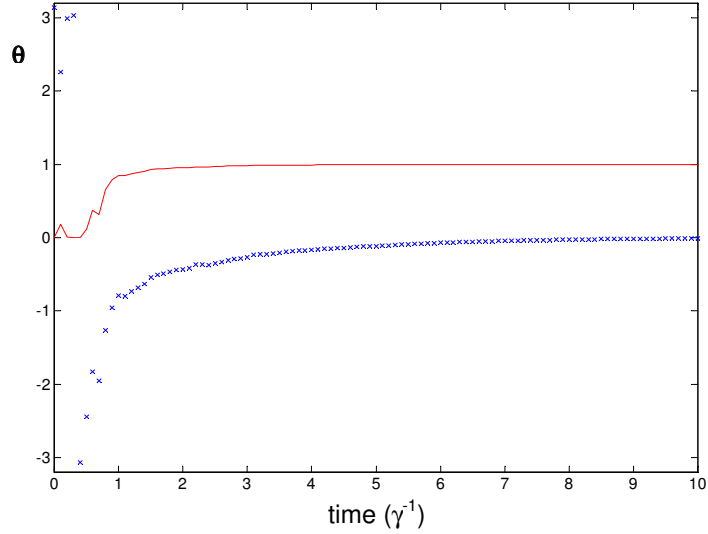}}
\caption{The two-level atom is driven from the ground state, $|1\rangle\equiv\hat{|\pi\rangle}$, to the excited state, $|2\rangle\equiv\hat{|0\rangle}$. The single quantum trajectories (crossed points) shown in (a) and (b) represent the two different types of paths that the state of the atom may follow. The two paths correspond to opposite directions around the Bloch sphere, passing through the (a) symmetric, or (b) anti-symmetric  superposition of the ground and excited states. The fidelity of $\hat{|\theta\rangle}$ compared to the target state $\hat{|0\rangle}$ is also shown (solid line) on the graphs. The transition time, $t_0$, defined here as the time taken to reach a state fidelity $|\langle\hat{\theta}|\hat{0}\rangle|^2\geq 0.99$ is (a) $t_0=4.0\gamma^{-1}$ and (b) $t_0=3.6\gamma^{-1}$. }
\label{Fig1}
\end{figure}

We have verified that the quantum feedback scheme works as intended, in the regime of unit detection efficiency and zero feedback delay, by simulating typical quantum trajectories for the feedback master equation in Eq. (\ref{me_feedback}). The transition time, $t_0$, required to drive the atom from an initial state to the target state, in the $x-z$ plane, is roughly proportional to the distance between these states, as measured by the difference in the angle $\theta$. The longest single-transition times $(\approx (4\pm 1)\gamma^{-1})$ correspond to driving the atom between two states that are diametrically opposite on the Bloch sphere, e.g., from the ground state to the excited state, as shown in Fig. \ref{Fig1}.

\section{Results and Comparison}
\label{Results and Comparison}

We first summarize the results of our open-loop control scheme, and then, whenever warranted, present a comparison between our results and the results of the quantum feedback scheme proposed by Wang {\it et al.} \cite{Wang01} for a two-level atom.

\subsection{Results for Open-loop Control}
\label{Results for Open-loop Control}

\begin{figure}[tp]
\scalebox{0.5}{\includegraphics{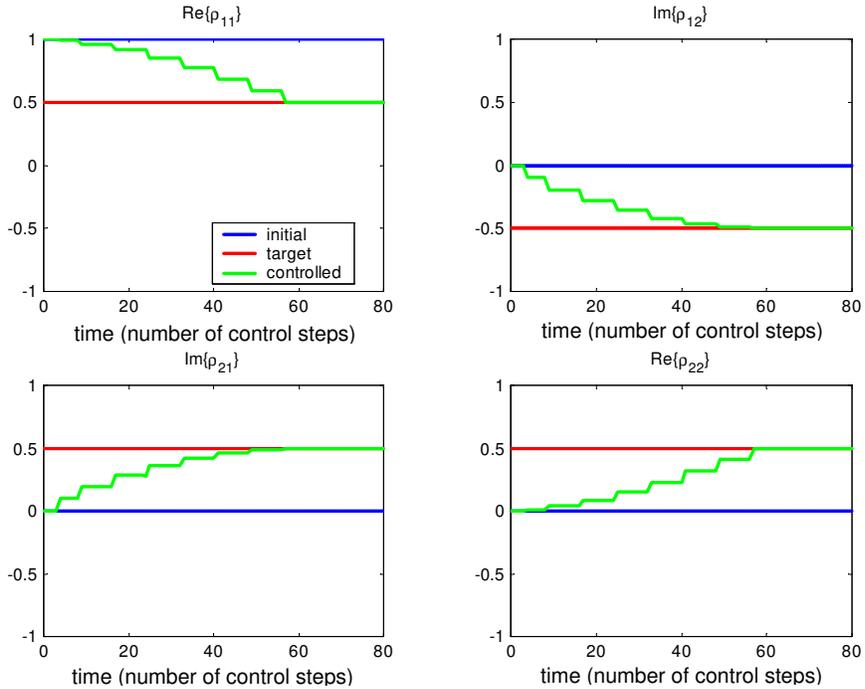}}
\caption{The non-zero density matrix elements of a qubit driven from the ground state, $|1\rangle$, to the symmetric superposition state, $(|1\rangle+i|2\rangle)/\sqrt{2}$. All of the density matrix elements show a monotonic transition to their target values. This driving used 6 intermediate states, with only one control cycle allocated for each intermediate state, and the maximum allowed strength of each control pulse was set by $I_{max}=0.1$. Note that the elements change in a noticeably discrete way during the transition period, and the fidelity of the final state is greater than $0.99$. }
\label{Fig2}
\end{figure}

First, we present the open-loop results for the evolution in the adiabatic decoherence regime, which corresponds to phase decay of the two-level system. Using a control Hamiltonian proportional to $\sigma_x$, the qubit can be driven to any target state on the surface of the Bloch sphere, in the $y-z$ plane. Only a small number of intermediate states are required to obtain robust control. The graphs in Fig. 2 illustrate the driving of the qubit from the ground state $|1\rangle$ to the symmetric superposition state $(|1\rangle+i|2\rangle)/\sqrt{2}$. The qubit was then driven to the excited state, and back to the ground state via the conjugate anti-symmetric superposition $(|1\rangle-i|2\rangle)/\sqrt{2}$, to complete a continuous circle around the Bloch sphere.

It is also possible to jump directly from the initial to the target state in a transition time as small as one control cycle, if the control pulses are strong enough. However, this can break the assumption that the magnitude of the control Hamiltonian is small, and moreover gradual driving improves the smoothness of the transition, albeit at the expense of longer transition times.

We have not simulated the control for the adiabatic decoherence regime using other control Hamiltonians, as there are no analogous results for the quantum feedback control of phase decay, either by itself, or in conjunction with spontaneous emission.

Next, we present the open-loop results for the evolution in the thermal decoherence regime, when the control Hamiltonian is proportional to $\sigma_y$. This corresponds to the ensemble evolution of the spontaneously-emitting two-level atom (without feedback) in the quantum feedback scheme. The numerical simulations used to implement our control strategy from Section 3, have assumed the ideal scenario: the decoherence function is known exactly, and the time step between control pulses for the open-loop control is small relative to the decoherence rate.

The graphs in Fig. 3 illustrate the driving of the qubit from the ground state $|1\rangle$ to the excited state $|2\rangle$ via the symmetric superposition $(|1\rangle+|2\rangle)/\sqrt{2}$, cf., Fig. 1(a) for the same driving in the quantum feedback scheme. One hundred intermediate states were used in this simulation to give a very smooth transition in a time $t_0 \approx 1250\Delta t$, where $\Delta t$ is the time interval between control pulses. The qubit was then driven back to the ground state via the anti-symmetric superposition $(|1\rangle-|2\rangle)/\sqrt{2}$, to complete a full circle around the Bloch sphere. We note that returning to the ground state with exactly the same phase, actually requires two complete circles around the Bloch sphere.

\begin{figure}[tp]
\scalebox{0.5}{\includegraphics{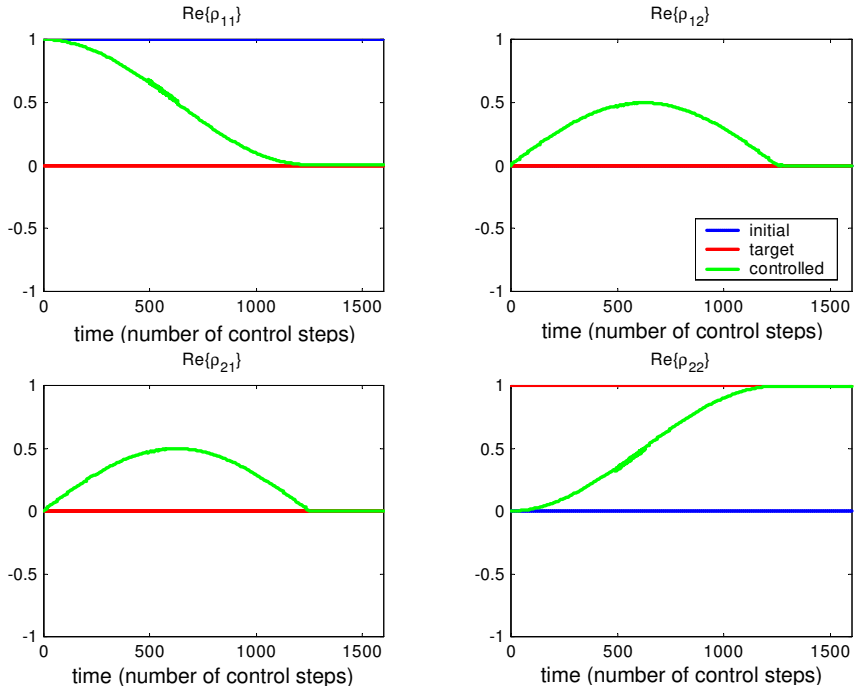}}
\caption{The non-zero density matrix elements of a qubit driven from the ground state, $|1\rangle$, to the excited state, $|2\rangle$. The population elements, $Re\{\rho_{11}\}$ and $Re\{\rho_{22}\}$, show a monotonic transition, whereas the coherences, $Re\{\rho_{12}\}$ and $Re\{\rho_{21}\}$, increase to $\frac{1}{2}$ as they approach the superposition state on the equator of the Bloch sphere, and then decrease back to zero. This driving used 100 intermediate states, with only two control cycles allocated for each intermediate state, and the maximum allowed strength of each control pulse was set by $I_{max}=0.01$. All of the density matrix elements change smoothly in the transition period, and the fidelity of the final state is greater than $0.99$. }
\label{Fig3}
\end{figure}

In general, our simulations of open-loop control in the thermal decoherence regime, show that the qubit can be driven to any pure state in the $x-z$ plane of the Bloch sphere, using a control Hamiltonian proportional to $\sigma_y$. Inevitably, the fidelity of the target state thus achieved fluctuates with each control cycle, but it stays close to one, given that the maximum allowed strength of the control pulses is relatively small.

The smoothness and length of the transition to the target state is determined by the number of intermediate states, as well as the maximum allowed strength of the control pulses. The graphs in Fig. 4 illustrate explicitly the same driving of the qubit as in Fig. 3, but using only ten intermediate states, and allowing the control pulses to be up to ten times stronger than before. We note that in the thermal decoherence regime, there is an absolute minimum number of intermediate states, of the order of ten states, to ensure successful control of the qubit, for the longest type of single-transition.

\begin{figure}[tp]
\scalebox{0.5}{\includegraphics{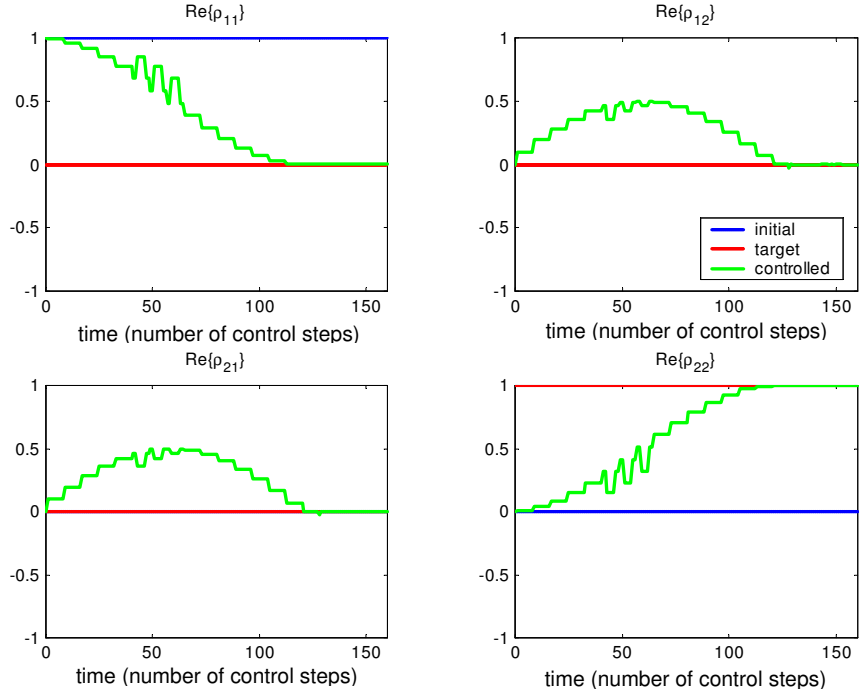}}
\caption{The non-zero density matrix elements of a qubit driven from the ground state, $|1\rangle$, to the excited state, $|2\rangle$. Only 10 intermediate states were used for this driving, with two control cycles allocated for each intermediate state, and the maximum allowed strength of each control pulse was set by $I_{max}=0.1$. All of the (non-zero) density matrix elements display large fluctuations in the transition period, and the fidelity of the final state is less than $0.99$.  }
\label{Fig4}
\end{figure}

The transition time from the initial state to the target state depends not only on the number of intermediate states, and the maximum allowable strength of the control pulses, but also on the magnitude of the time step, $\Delta t$, between control pulses. The minimum times we observed for the longest single-transition of the qubit e.g., from the ground state to the excited state, are of the order of ten control cycles, for the type of low-quality results shown in Fig. 4. Of course, the transition times grow longer as the performance of the open-loop scheme improves, e.g., the driving shown in Fig. 3 requires a transition time of the order of 100-200 control cycles.

We noted earlier, that our control Hamiltonian, and the equivalent interaction Hamiltonian due to the classical driving field in the quantum feedback scheme, can be taken as a linear superposition of $\sigma_x$ and $\sigma_y$, see Eq. (\ref{H_Ic}). In general, if the control Hamiltonian is written in terms of the angle $\phi$,
\begin{equation}
H_{Ic}(\phi) \propto (\sigma_x \sin\phi +  \sigma_y \cos\phi),
\label{H_phi}
\end{equation}
the stationary solutions of the evolution,
\begin{eqnarray}
x_{ss} &=& \frac{4\alpha\gamma\cos\phi}{\gamma^2+8\alpha^2} \\
y_{ss} &=& \frac{-4\alpha\gamma\sin\phi}{\gamma^2+8\alpha^2} \\
z_{ss} &=& \frac{-\gamma^2}{\gamma^2+8\alpha^2},
\end{eqnarray}
are contained in a plane $S_\phi$, and lie on a curve parametrized by the ratio, $\alpha/\gamma$, of the driving field and the decoherence rate.

The plane $S_\phi$, defined by $x\sin\phi+y\cos\phi=0$, always contains the $z$-axis (i.e., always includes both the ground and the excited states of the atom), and is rotated by an angle $\phi$ around the $z$-axis in Bloch space, with respect to the reference $x-z$ plane $(S_{\phi=0})$. Thus the effect of the driving field corresponding to $H_{Ic}(\phi)$ is to rotate the state of the atom around an axis which is normal to the plane $S_\phi$.

It is evident that if we employ different control Hamiltonians, $H_{Ic}(\phi)$, the stationary solutions for the evolution of the spontaneously-emitting two-level atom (without feedback) are not restricted to the $x-z$ plane. This result also applies to the quantum feedback scheme of Wang \textit{et al.} and our open-loop control in the adiabatic case, enabling us to stabilize a qubit in any pure state (except for equal superpositions) on the surface of the Bloch sphere.

In general, the atom can be driven reversibly to any arbitrary pure target state on the Bloch sphere, by a combination of two control Hamiltonians. The first Hamiltonian rotates the initial state to the ground state, and the second Hamiltonian rotates the ground state to the desired target state. To return to the initial state, the order and the sign of the Hamiltonians need to be reversed.

To demonstrate this idea, we have also simulated driving the qubit in the thermal decoherence regime, using a control Hamiltonian proportional to $\sigma_x$. The same principle applies if the control Hamiltonian is a linear superposition of $\sigma_x$ and $\sigma_y$, though the computation of the evolution operator $U_I$ and the density matrix elements is more intensive.

\begin{figure}[tp]
\scalebox{0.5}{\includegraphics{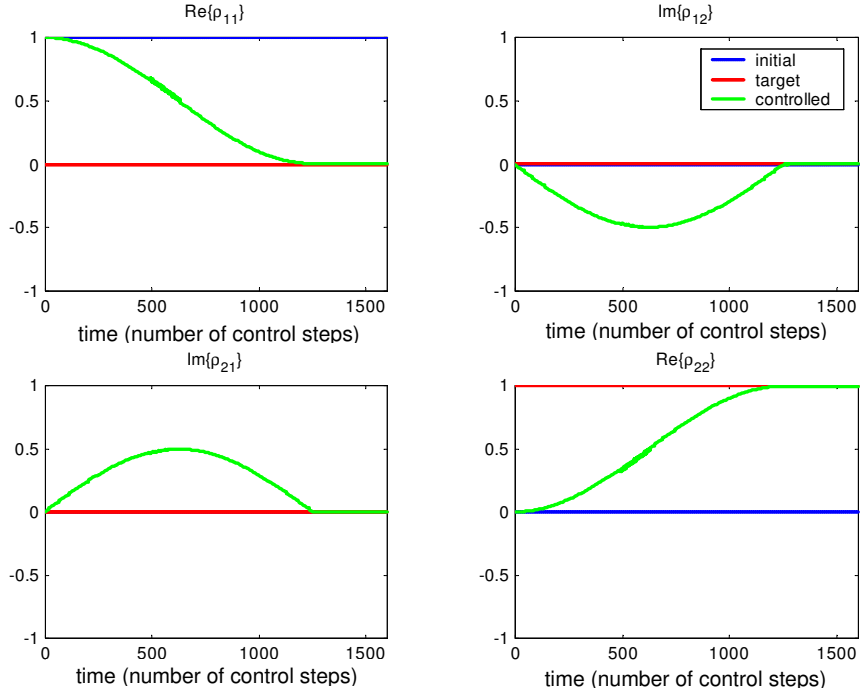}}
\caption{The non-zero density matrix elements of a qubit driven from the ground state, $|1\rangle$, to the excited state, $|2\rangle$. The population elements, $Re\{\rho_{11}\}$ and $Re\{\rho_{22}\}$, show a monotonic transition, whereas the coherences, $Im\{\rho_{12}\}$ and $Im\{\rho_{21}\}$, increase to $\frac{1}{2}$ as they approach the superposition state on the equator of the Bloch sphere, and then decrease back to zero. This driving used 100 intermediate states, with only two control cycles allocated for each intermediate state, and the maximum allowed strength of each control pulse was set by $I_{max}=0.01$. All of the density matrix elements change smoothly in the transition period, and the fidelity of the final state is greater than $0.99$. }
\label{Fig5}
\end{figure}

The graphs in Fig. 5 illustrate explicitly the driving of the qubit from the ground state $|1\rangle$ to the excited state $|2\rangle$, using a control Hamiltonian $H^\prime_{Ic}\propto\sigma_x$, c.f., Fig. 3 for the same driving with $H_{Ic}\propto\sigma_y$. The state of the qubit is confined to the $y-z$ plane of the Bloch sphere (and passes through the symmetric superposition state $(|1\rangle+i|2\rangle)/\sqrt{2}$), thus the effect of the driving field is to rotate the state of the atom around the $x$-axis. One hundred intermediate states were used in this simulation to give a very smooth transition in a time $t_0 \approx 1250\Delta t$, where $\Delta t$ is the time interval between control pulses. The qubit was then driven back to the ground state via the conjugate anti-symmetric superposition $(|1\rangle-i|2\rangle)/\sqrt{2}$, to complete a continuous circle around the Bloch sphere.

\begin{figure}[tp]
\scalebox{0.6}{\includegraphics{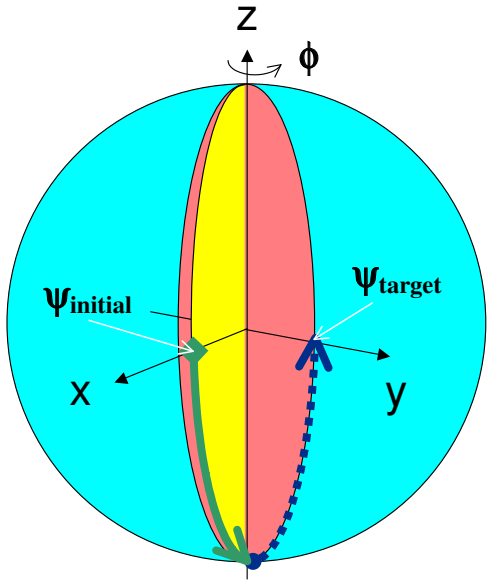}}
\caption{The effect of two orthogonal control Hamiltonians for pure states on the surface of the Bloch sphere. The qubit is driven from the initial state $\psi_{initial}$ to the ground state $|1\rangle$ (solid arrow), using a control Hamiltonian $H_{Ic}\propto\sigma_y$. Next, a control Hamiltonian $H_{Ic}\propto\sigma_x$ is used to drive the qubit from the ground state to the target state $\psi_{target}$ (broken arrow). }
\label{Fig6}
\end{figure}

To illustrate how a combination of two control Hamiltonians can be used to achieve any arbitrary pure target state, suppose that we are given the initial state $\psi_{initial}=(|1\rangle+|2\rangle)/\sqrt{2}$, in the $x-z$ plane, as shown in Fig. \ref{Fig6}. If we are asked to drive the qubit to the target state $\psi_{target}=(|1\rangle+i|2\rangle)/\sqrt{2}$, in the $y-z$ plane, our control strategy is as follows: first use the control Hamiltonian $H_{Ic}\propto\sigma_y$ to drive the qubit to the ground state, along the curve determined by the intersection of the Bloch sphere with the $x-z$ plane, and then use the control Hamiltonian $H^\prime_{Ic}\propto\sigma_x$ to drive the qubit to the target state, along the curve determined by the intersection of the Bloch sphere with the $y-z$ plane.

\subsection{Comparison between the Open-loop and Quantum Feedback Controls}
\label{Comparison between the Open-loop and Quantum Feedback Controls}

The open-loop evolution in the thermal decoherence regime, is identical to the ensemble evolution of the spontaneously-emitting two-level atom (without feedback) modeled by Wang \textit{et al.} \cite{Wang01}, when the control Hamiltonian is proportional to $\sigma_y$. After quantum feedback is introduced, the difference in the dynamics between the two schemes becomes evident: the feedback directly changes the effective interaction of the atom with its environment, whereas in the open-loop scheme, the control Hamiltonian, which is equivalent to the classical driving field $\alpha$ in Eq. (\ref{qf_H_I}), affects only the atom.

Both the open-loop and the quantum feedback control schemes presented in this paper aim to drive a qubit to an arbitrary target state, and then maintain that state. For simplicity, we have restricted our attention to pure states, though in principle the initial and target states may be mixed states in both schemes. The general case will be investigated in future work to check our conjecture that it is only possible to drive to a target state that has a purity less than or equal the purity of the initial state.

The initial conditions are not important for the quantum feedback scheme, whereas with open-loop control we must assume prior knowledge of the initial state. This requirement can be relaxed for our scheme, if we are able to reset all initial states to the ground state, so that we have a fixed starting point for the qubit. Finally, neither scheme can be used to maintain an \textit{unknown} state such as that obtained from a quantum computation. Both schemes can only drive the atom to a \textit{known} target state.

If the evolution in our model is the same as the ensemble evolution in the quantum feedback scheme, the open-loop control can drive the qubit between any two pure states in the $x-z$ plane (see the results illustrated in Fig. 3). Therefore, our simulations have demonstrated that open-loop control can effectively achieve the same result as the quantum feedback scheme.

We discuss in turn the (i) efficiency, (ii) feasibility, and (iii) robustness of the open-loop and quantum feedback schemes:

\subsubsection{Efficiency.}
The final state actually achieved by the open-loop control scheme has a fidelity which fluctuates during each control cycle, but remains very close to one with respect to the target state, given that the maximum allowed strength of the control pulses is relatively small. The transition times in our scheme are much shorter than the equivalent times for the quantum feedback scheme, if the rate of the control pulses is higher than $\approx 1000\gamma$.

\subsubsection{Feasibility.}
The quantum feedback scheme relies on the assumptions of perfect detection and zero delay time. In the case of inefficient detection, or in the non-Markovian regime (non-zero delay), the final state is a mixed state \cite{Wang01}. The deviation from a pure target state is noticeable even for relatively small violations e.g., $95\%$-efficiency or $\tau_F\geq0.02\gamma^{-1}$, especially for states close to the equal superposition states at the equator of the Bloch sphere, which are inherently unstable. In terms of practical implementation, it is currently impossible to satisfy these assumptions due to technological challenges.

On the other hand, the requirements for the open-loop control scheme are in principle less stringent. The time step between control pulses needs to be small relative to the decoherence rate, which can be satisfied in a practical implementation, by choosing an atomic transition with a decoherence rate that is low compared to available control pulse rates. The strength of the control pulses and the strength of the system-environment interaction also need to be small relative to the free Hamiltonian, in order to validate the first-order approximation of the evolution operator obtained in Eqs.(\ref{U_I_a}) and (\ref{U_I_t}). The other major assumption is that the decoherence function is known accurately, which is essentially true for a single two-level system such as a trapped atom.

\subsubsection{Robustness.}
The stochastic model used in the quantum feedback scheme describes the evoulution of single quantum systems using quantum trajectories with an extrinsic noise source. On the other hand, the spin-boson model assumes knowledge of the average decoherence function, and a deterministic evolution, which corresponds to an ensemble of quantum systems.

The stochastic model is a more realistic portrayal of the quantum dynamics of a single spontaneously-emitting two-level atom than the generic ensemble model we have employed. Thus, in principle, the quantum feedback scheme is always going to perform more robustly than any open-loop scheme, because it is inherently able to use feedback to control a single qubit. Following the transition period, the effect of random fluctuations on the state of a single qubit are effectively canceled by the feedback.

The spin-boson model uses an average decoherence function describing the behavior of an ensemble of qubits. Our scheme cannot respond adaptively to random fluctuations for a single qubit, which will result in much greater deviations from the target state, and the open-loop control may fail completely, if the fluctuations are too large.

\vspace{20pt}
In conclusion, we presented an open-loop control scheme that drives a qubit to an arbitrary target state with high fidelity, while maintaining quantum coherence throughout the process.

We have also shown that by using different control Hamiltonians, which are proportional to a linear combination of $\sigma_x$ and $\sigma_y$, the quantum feedback scheme of Wang \textit{et al.} can be extended to stabilize the qubit in any known pure state, except for equal superposition states.

The performance of our scheme compares favorably with the performance of the quantum feedback scheme proposed by Wang \textit{et al.}, with respect to target fidelity, and transition time. In principle, both schemes can drive a qubit to an arbitrary pure target state, with high fidelity, in a time of the order of $10\gamma^{-1}$ or less. 

From a practical viewpoint though, while the quantum feedback scheme has the capability to react to sudden fluctuations for a single qubit, its performance suffers from the strict technical requirements for zero feedback delay and unit detection efficiency. The open-loop control scheme has the additional advantage that the transition time can be lowered by simply increasing the rate of the control pulses.

\section* {Acknowledgments}
\label {Acknowledgments}
{This research was supported in part by the U.S. Department of Energy, Office of Basic Energy Sciences.  The Oak Ridge National Laboratory is managed for the U.S. DOE by UT-Battelle, LLC, under contract No. DE-AC05-00OR22725. C.D. thanks Howard Wiseman for his explanations regarding the quantum feedback scheme.}


\begin{thebibliography}{99}
\bibitem{Viola98}Viola L and Lloyd S 1998 Phys. Rev. A {\bf 58}(4) 2733
\bibitem{Viola99a}Viola L, Knill E and Lloyd S 1999 Phys. Rev. Lett. {\bf 82}(12), 2417
\bibitem{Viola99b}Viola L, Lloyd S and Knill E 1999 Phys. Rev. Lett. {\bf 83}(23), 4888
\bibitem{Tombesi_ol}Vitali D and Tombesi P 1999 Phys. Rev. A {\bf 59} 4178; Vitali D and Tombesi P 2002 Phys. Rev. A {\bf 65} 012305; Mancini S, Vitali D, Bonifacio R and Tombesi P 2001 {\it Preprint} quant-ph/0108011
\bibitem{Zanardi97}Zanardi P and Rasetti M 1997 Phys. Rev. Lett. {\bf 79}, 3306
\bibitem{Lidar98}Lidar D A, Chuang I L and Whaley K B 1999 Phys. Rev. Lett. {\bf 81}, 2594
\bibitem{Preskill98}Preskill J 1998 Proc. Roy. Soc. Lond. A {\bf 454}, 385
\bibitem{Knill00}Knill E, Laflamme R and Viola L 2000 Phys. Rev. Lett. {\bf 84}, 2525
\bibitem{NC} Nielsen M A and Chuang I L 2000 {\it Quantum Computation and Quantum Information} (Cambridge: Cambridge University Press)
\bibitem{Wang01}Wang J and Wiseman H 2001 Phys. Rev. A {\bf 64} 063810;  Wang J, Wiseman H and Milburn G J 2001 J. Chem. Phys. {\bf 268} 221
\bibitem{Tombesi_qf}Tombesi P and Vitali D 1995 Phys. Rev. A {\bf 51} 4913; Goetsch P, Tombesi P and Vitali D 1996 Phys. Rev. A {\bf 54} 4519; Tombesi P and Vitali D 1995 Appl. Phys. B {\bf 60} S69; Tombesi P, Vitali D and Milburn G J 1997 Phys. Rev. Lett. {\bf 79} 2442; Tombesi P, Vitali D and Milburn G J 1998 Phys. Rev. A {\bf 57} 4930; Fortunato M, Raimond J M, Tombesi P and Vitali D 1999 Phys. Rev. A {\bf 60}, 1687
\bibitem{Protopopescu02}Protopopescu V, Perez R B, D'Helon C and Schmulen J 2003 J. Phys. A {\bf 36} 2175
\bibitem{Khaneja_control} Khaneja N, Brockett R and Glaser S J 2001 Phys. Rev. A {\bf 63} \#032308; Khaneja N and Glaser S J 2001 Chem. Phys. {\bf 267} 11; Khaneja N, Glaser S J and Brockett R 2002 Phys. Rev. A {\bf 65} \#032301; Khaneja N and Glaser S J 2002 Phys. Rev. A {\bf 66} \#060301
\bibitem{Khaneja_relax} Khaneja N, Reiss T, Luy B and Glaser S J 2002 {\it Preprint} quant-ph/0208050; Khaneja N, Luy B and Glaser S J 2003 {\it Preprint} quant-ph/0302060
\bibitem{Weiss99}Weiss U 1999 {\it Quantum Dissipative systems} (Singapore: World Scientific)
\bibitem{Scully_qo} Scully M S and Zubairy M S 1996 {\it Quantum Optics} (Cambridge: Cambridge University Press)
\bibitem{Gardiner_qn} Gardiner C W and Zoller P 2000 {\it Quantum Noise} (Berlin: Springer)
\bibitem{Palma96}Palma G M, Suominen K -A and Ekert A K 1996 Proc. R. Soc. London A {\bf 452} 567
\bibitem{Wiseman_feedback} Wiseman H M and Milburn G J 1993 Phys. Rev. Lett. {\bf 70}, 548; Wiseman H M 1994 Phys. Rev. A {\bf 49}, 2133; Wiseman H M 1994 Phys. Rev. A {\bf 49}, 1350
\bibitem{Milburn_qo} Walls D F and Milburn G J 1994 {\it Quantum Optics} (Berlin: Springer)
\end{thebibliography}
\end{document}